\documentclass[aip,graphicx,reprint,amsmath,amssymb]{revtex4-1}

\usepackage{graphicx}
\usepackage{siunitx}
\usepackage[version=3]{mhchem}
\usepackage{amsmath}
\usepackage{amssymb}
\usepackage{natmove}

\begin{document}

\title{Metadynamics for Automatic Sampling of Quantum Property Manifolds: Exploration of Molecular Biradicality Landscapes}

\author{Joachim O. Lindner}
\affiliation{Center for Nanosystems Chemistry (CNC), Julius-Maximilians-Universit\"{a}t W\"{u}rzburg, Theodor-Boveri-Weg,
97074 W\"{u}rzburg, Germany}
\affiliation{Institut f\"{u}r Physikalische und Theoretische Chemie,
Julius-Maximilians-Universit\"{a}t W\"{u}rzburg,
Emil-Fischer-Str. 42, 97074 W\"{u}rzburg, Germany}
\author{Merle I. S. R\"ohr}
\email{merle.roehr@uni-wuerzburg.de}
\affiliation{Center for Nanosystems Chemistry (CNC), Julius-Maximilians-Universit\"{a}t W\"{u}rzburg, Theodor-Boveri-Weg,
97074 W\"{u}rzburg, Germany}

\date{\today}

\begin{abstract}
We present a general extension of the metadynamics allowing for an automatic sampling of quantum property manifolds (ASQPM) giving rise to functional landscapes that are analogous to the potential energy surfaces in the frame of the Born-Oppenheimer approximation. For this purpose, we employ generalized electronic collective variables to carry out biased molecular dynamics simulations in the framework of quantum chemical methods that explore the desired property manifold. We illustrate our method on the example of the "biradicality landscapes",  which we explore by introducing the natural orbital occupation numbers (NOONs) as the electronic collective variable driving the dynamics. We demonstrate the applicability of the method on the simulation of \textit{p}-xylylene and [8]annulene allowing to automatically extract the biradical geometries. In the case of [8]annulene the ASQPM metadynamics leads to the prediction of biradical scaffolds that can be stabilized by a suitable chemical substitution, leading to the design of novel functional molecules exhibiting biradical functionality.
\end{abstract}


\maketitle

\section{Introduction}
 Molecular functionality can often be assigned to a given property of the electronic wavefunction, e.g., the strength of the transition dipole moment determines the efficiency in light-harvesting processes,\cite{oviedo2011} while in catalytic processes, properties such as the charge distribution and spin-density define the intrinsic activity of the catalysts as well as their reaction route.{\cite{galvan1986,Remenyi2005,NaYang2018}} Since the electronic wavefunction exhibits a direct dependence on the nuclear coordinates of the molecule, its characteristics are directly dependent on the structural evolution of the system. Therefore, within the framework of the Born-Oppenheimer (BO) approximation\cite{bornoppenheimer1927}, these properties can be represented as "functional landscapes", in analogy to the concept of potential energy surfaces (PES). The theoretical investigation of these multidimensional hypersurfaces enables a systematic exploration of the structure-function relationship in molecules, allowing to obtain information about their stability and thus to ascertain the structural and energetic accessibility of desired functions in molecules and functional conformations.  
 A molecular functionality of particular interest consists in the biradical character.\cite{Abe2013} Several light-induced processes in molecules take place upon transient population of biradical conformations, e. g. at crossing points of the first excited and the ground BO states, where strong nonadiabatic coupling leads to efficient non-radiative decay of excited molecules.\cite{Bonacic-Koutecky1987} Also several bond forming reactions take place via biradical intermediates, such as photoinduced cycloaddition reactions.\cite{Alcaide2010}
 The introduction of protecting groups allows to stabilize these highly reactive compounds, giving rise to new classes of functional molecules with several potential applications:
 Organic biradical molecules serve as attractive candidates for optoelectronic applications such as singlet fission \cite{Nakano,Smith_2010,Wu2015}. Successful strategies for the realization of new organic biradicals consist in consideration of quinoidal compounds and to produce their open-shell resonance structure upon substitution, following the ideas of i) strengthening the $\pi$ electron conjugation and thus aromatic structure or ii) biasing the biradical configuration by introduction of a steric strain.\cite{Defrancisco_2018}
 Also inorganic biradicals can be stabilized upon introduction of protecting ligands. Only recently, a biradical complex bearing a twisted boron-boron double bond stabilized by the introduction of bulky cyclic (alkyl)(amino)carbene (CAAC) ligands has been presented.\cite{Boehnke2018}

Therefore, development of theoretical methods that allow for systematic generation of new biradical molecular scaffolds is highly desired. Accompanied with the development of strategies for the energetic stabilization of the generated biradicals this would enable systematic prediction of new molecules with desired biradical-based functionality.

In this contribution we wish to present a new methodology for an "automatic sampling of quantum property manifolds" (ASQPM) which allows to theoretically investigate the functional landscape of molecules with a given chemical composition. Therefore, based on the idea of Parrinello's metadynamics \cite{Laio2002,Barducci2008,Raiteri_2006,Bonomi2010,Tribello2010,Branduardi2012} we have developed a multistate extension of the latter by introduction of quantum mechanical, electronic collective variables that are capable to represent the function of interest. Furthermore, we employ an additional bias that serves to locally modify the already explored regions, driving the dynamics forward, in the limit sampling the whole conformational space. 

As a first demonstration, we apply this method in the framework of the complete active space self-consistent field method (CASSCF) to the exploration of "biradicality landscapes" of \textit{p}-xylylene and [8]annulene.  From the  theoretical point of view, biradicality can be clearly definded employing the occupation numbers of natural orbitals (NOONs) which can be derived from the electronic wavefunction upon calculation and subsequent diagonalization of the density matrix. Therefore, sampling the NOONs within the structural conformation space gives a direct picture of the "biradicality landscape" as presented for \textit{p}-xylylene.
In addition, we have carried out further calculations on the biradical conformations of [8]annulene obtained from the ASQPM simulations, enabling the design of stable biradical molecules upon rational substitution, proving that the ASQPM simulations can deliver useful information for the realization of new functional molecules.
 
\section{Method}
 
The original metadynamics introduced by Parrinello and coworkers \cite{Laio2002,Barducci2008,Raiteri_2006,Bonomi2010,Tribello2010,Branduardi2012} represents a sophisticated accelerated molecular dynamics technique using collective variables (CVs) to drive transitions between different barrier-separated basins on  the  PES,  thus allowing  for  systematic  sampling  of  the  PES
as  well  as the determination  of  free  energies. A  wide  variety  of classic CVs  has  been
implemented, ranging from simple geometrical quantities to complex
variables constructed by machine learning and dimensionality
reduction techniques.\cite{Tribello2010, Ceriotti2011} The method finds broad application in the frame of force field simulations e.g. for investigation of folding mechanisms in proteins,\cite{Piana2007} while studies based on quantum chemical metadynamics simulations are still relatively rare.\cite{Grimme2019,Luber2017}

To the best of our knowledge, we have recently published a first multistate formulation of the quantum chemical metadynamics, capable of the automatic localization of conical intersections in molecular systems. The algorithm is available through the metaFALCON program package \cite{Lindner2018,Lindner2019}. 
In the present study, we employ the NOONs as an electronic collective variable, which can be derived from the electronic wavefunction by calculating the one-electron reduced density matrix expanded into the molecular orbital basis. Upon diagonalization, a set of eigenvalue-eigenvector pairs is obtained, representing the reduced density matrix $\mathbf{D}$ in terms of natural orbitals (NO) $\varphi_i$ as
\begin{equation}
    \mathbf{D}(\mathbf{x}_1, \mathbf{x}_1') = \sum_{i} n_i \varphi_i^*(\mathbf{x}_1) \varphi_i(\mathbf{x}_1').
\end{equation} 
It should be noticed that the NOs are the best possible approximation of the many-particle problem based on one-electron wavefunctions and can therefore be used as an indicator for unpaired electrons in a molecular system.  The corresponding eigenvalues $n_i$ can be interpreted as NOONs \cite{Flynn1974,Doehnert1980}.

In a closed-shell configuration, all NOONs lie either close to zero or two, i.e. electrons occupy the NOs in pairwise manner. The orbital with the lowest NOON close to two is referred to as highest occupied natural orbital (HONO) with NOON $n_H$, while the orbital with the highest NOON close to zero is called lowest unoccupied natural orbital (LUNO) with NOON $n_L$. In an open-shell configuration with even total number of electrons, however, one or several pairs of NOs are singly occupied leading to NOON values close to 1. In the special case of a biradical, two unpaired weakly interacting electrons are present, i.e. $N/2-1$ NOs are doubly occupied, followed by the two NOs with NOON value $n_H$ and $n_L$ that are approximately one. Therefore, the gap between $n_H$ and $n_L$  can serve as a measure for the degree of biradicality (if all other NOONs have values close to zero or two), and thus can be employed as a clearly defined, quantum-mechanical collective variable for the systematic sampling of the biradicality landscape.

\begin{figure}
\begin{centering}
\includegraphics{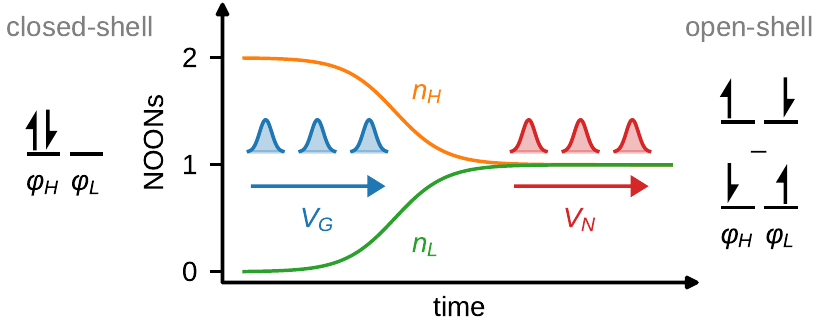}
\par\end{centering}
\protect\caption{Scheme illustrating the ASQPM algorithm for exploration of the biradicality landscape. At $t=0$ the system
has a closed shell electronic structure with $n_H=2$ and $n_L=0$ indicated also by the closed shell electronic configuration in the left part of the figure. Periodic addition of the biasing potentials $V_G$ along the $\Delta n$ collective variable leads to the gap closing, which is associated with the formation of a singlet biradical geometry characterized by $n_H=1$ and $n_L=1$. The exploration of the biradicality landscape is enforced by a periodic addition of an off-diagonal bias $V_N$ forcing the system to explore different geometries with an open shell biradical structure indicated by the electronic configuration in the right part of the figure.\label{fig:scheme}}
\end{figure}

Our ASQPM method relies on a molecular dynamics simulation propagating Newtons equation of motion, biased by the modification of the potential energy surface (PES) upon adding a history-dependent potential $V_{G}(t)$:
\begin{equation}
m_i \ddot{\mathbf{R}}_i=-\nabla _i(V_{el}+V_{G}(t)).\label{eq:mdforces}
\end{equation}
with $V_{el}$ representing the unbiased BO PES.
The bias potential $V_{G}(t)$ is built up at regular time steps $\tau_{G}$ by adding gaussian-shaped functions along the electronic CV $\Delta n_{meta}$ :
\begin{equation}
\begin{aligned}
    V_{G}(t)= & \sum_{t'=\tau_{G},2\tau_{G},\ldots}^{t} w\exp\left(-\frac{\left(\Delta n_{meta}(t)-\Delta n_{meta}(t')\right)^{2}}{2{\delta \Delta n_{meta}}^{2}}\right) \\
    & \times\Theta(\Delta n_{meta}(t')-\epsilon),\label{eq:metadynamics}
\end{aligned}
\end{equation}

 while the shape of the added Gaussians is defined by the height $w$ and width $\delta \Delta n_{meta}$ parameters. The Heaviside $\Theta$-function is used in order to update the potential only when $\Delta n_{meta}$ exceeds the threshold $\epsilon$. In order to systematically sample the biradicality landscape we introduce an additional off-diagonal coupling term $V_{N}$ into the sub-block of the density matrix $\mathbf{D_{diag}}$ containing the HONO and LUNO (represented by their NOON values $n_{H}$ and $n_{L}$). In this way, the corresponding submatrix becomes
\begin{equation}
    \mathbf{D_{meta}} =
    \begin{pmatrix}
        n_{L} & V_{N} \\
        V_{N} & n_{H}
    \end{pmatrix}.
\end{equation}
Upon diagonalization of the latter, we obtain an effective NOON gap
\begin{equation}
    \Delta n_{meta} = \sqrt{\left(n_{L}-n_{H}\right)^2+4V_{N}^2}.
\end{equation}
which assumes values between 2 (closed shell structure) and 0 (bi- or polyradical structure), as long as $V_{N}$ is zero. The off-diagonal biasing potential $V_{N}$ plays a crucial role in the method allowing the system to move along the biradicality landscape. For this purpose, $V_{N}$ is updated every $\tau_G$ steps, if $\Delta n_{meta}$ drops below $\epsilon$, according to
\begin{equation}
\begin{aligned}
    V_{N}(t)= & \sum_{t'=\tau_{G},2\tau_{G},\ldots}^{t}w_{bi}\exp\left(-\sum_{i}\frac{\left(s_{i}(t)-s_{i}(t')\right)^{2}}{2{\delta s_{i}}^{2}}\right) \\
    & \times\Theta(\epsilon-\Delta n_{meta}(t')).\label{eq:metadynamics_vnoon}
\end{aligned}
\end{equation}
In order to distinguish between different biradical geometries and prevent the simulation to visit the same part of the landscape several times, a set of additional (geometric) CVs $\{ s_{i} \}$ needs to be chosen.
The algorithm and the role of the biasing potentials is schematically illustrated in Fig. \ref{fig:scheme}.
The restriction to the exploration of biradical structures rather than polyradicals, i.e. all NOONs other than $n_{L}$ and $n_{H}$ are close to 0 or 2,  is achieved by two quadratic wall potentials acting on $n_{L+1}$ and $n_{H-1}$
\begin{equation}
    V_{wall}^{L+1} =
    \begin{cases}
    k(n_{L+1}-\epsilon_{L+1})^2 & \text{for $n_{L+1}>\epsilon_{L+1}$} \\
    0 & \text{for $n_{L+1}<\epsilon_{L+1}$}
    \end{cases},
    \label{eq:walll}
\end{equation}
\begin{equation}
    V_{wall}^{H-1} =
    \begin{cases}
    k(n_{H-1}-\epsilon_{H-1})^2 & \text{for $n_{H-1}<\epsilon_{H-1}$} \\
    0 & \text{for $n_{H-1}<\epsilon_{H-1}$}
    \end{cases}.
    \label{eq:wallh}
\end{equation}
Due to the sorting of the NOONs, these two wall potentials restrict all NOONs other than $n_H$ and $n_L$ to values below $\epsilon_{L+1}$ or above $\epsilon_{H-1}$.

Thus, the full equation of motion for the nuclei in the frame of the ASQPM method has the form:
\begin{equation}
    m_i \ddot{\mathbf{R}}_i=-\nabla _i(V_{el}+V_{G}+V_{wall}^{L+1}+V_{wall}^{H-1}),\label{eq:mdforces_asqpm}
\end{equation}
where $V_{el}$ is the original electronic potential obtained from the chosen electronic structure method and $V_{G}$ and $V_{wall}^{L+1}+V_{wall}^{H-1}$ represent the above described biasing potentials.

Evaluation of the nuclear gradients requires the differentiation of $\Delta n_{meta}$ with respect to the nuclear coordinates
\begin{equation}
    \nabla \left( \Delta n_{meta} \right) =
    \frac{(n_L-n_H) \nabla \left( n_L-n_H \right) + 4V_N \nabla V_N}
    {\Delta n_{meta}}.
\end{equation}
while the gradients $\nabla n_H$ and $\nabla n_L$ are needed for calculation of the wall potentials in Eqs.~\ref{eq:walll} and \ref{eq:wallh} and are available through numerical differentiation, which can be efficiently parallelized.

For initializing an ASQPM simulation any arbitrary structure of the studied molecular system can be used. 
For integration of the nuclear equations of motion we employ the velocity Verlet algorithm \cite{Swope1982}. The nuclear forces are calculated ''on the fly'' according to Eq.~\ref{eq:mdforces_asqpm}. If the starting structure favors a closed shell configuration, the dynamics is dominated by the electronic potential $V_{el}$ in the beginning. As the simulation proceeds, Gaussian-shaped potentials are added to $V_G$, until the influence of the latter is strong enough to lead the system to a biradical structure with $\Delta n_{meta} \approx 0$. From there on, $V_N$ has growing impact on the dynamics because the small $\Delta n_{meta}$ activates the addition of Gaussians according to Eq.~\ref{eq:metadynamics_vnoon}, thus starting the biradicality landscape exploration phase. The interplay of the two bias potentials $V_G$ and $V_N$ allows to subsequently visit multiple different biradical structures while $V_N$ converges to a hypersurface that can be interpreted as "biradical functional landscape" with respect to the (geometric) CVs $s_i$.

\section{Results and Discussion}
\subsection{\textit{p}-Xylylene}
\begin{figure}
\begin{centering}
\includegraphics{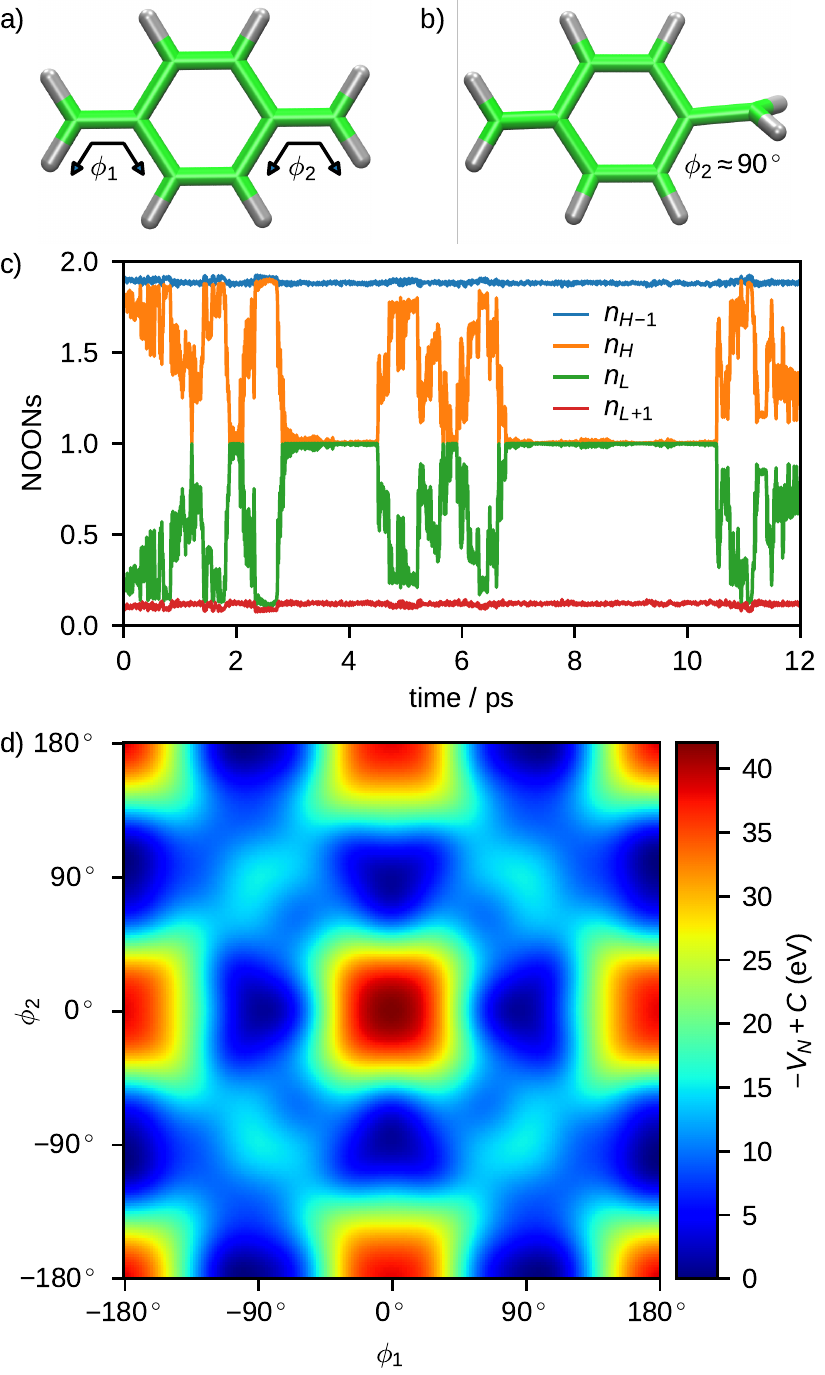}
\par\end{centering}
\protect\caption{a) Structure of \textit{p}-xylylene and definition of CVs $\phi_{1}$ and $\phi_{2}$. b) Biradical structure reached after \SI{1.2}{ps} with one of the methylene groups twisted by almost \ang{90}. c) Evolution of frontier NOONs along the ASQPM trajectory. d) Biradicality hypersurface constructed from the metadynamics potential $V_{N}$. The lowest point of the surface has been set to zero. \label{fig:xylylene}}
\end{figure}

In contrast to \textit{m}-xylylene which is a biradical in its most stable singlet ground state structure, a closed-shell quinoidal structure is preferred for \textit{p}-xylylene \cite{Schaefgen1955,Coulson1947,Williams1970}. However, rotation of one of the methylene groups by \ang{90} separates one electron at the respective carbon atom from the planar $\pi$-system and therefore changes the electronic character from closed-shell to biradical qualifying \textit{p}-xylylene as an illustrative example for the demonstration of NOON-gap metadynamics. We use the two torsion angles $\{ \phi_1, \phi_2 \}$ defining the orientation of the methylene groups with respect to the ring plane as a set of CVs in Eq.~\ref{eq:metadynamics_vnoon} (see Fig.~\ref{fig:xylylene}a), proving in the following section that the ASQPM approach is suitable to find the expected biradical conformations.

Starting the ASQPM simulations from planar \textit{p}-xylylene (shown in Fig.~\ref{fig:xylylene}a), the bias potential $V_G$ drives the molecular dynamics into a biradical configuration space within \SI{1.2}{ps}. The resulting structure, displayed in Fig.~\ref{fig:xylylene}b, exhibits a \ang{90} twist and  pyramidalization of the methylene group. The progression of the frontier NOONs along the \SI{12.0}{ps} trajectory presented in Fig.~\ref{fig:xylylene}c clearly indicates that the initial electronic structure is of closed-shell singlet character in the beginning, reflected in occupation numbers close to 0 and 2 while during the first \SI{1.2}{ps} the gap between $n_H$ and $n_L$ decreases, reaching an occupation of 1 and 1, respectively. After \SI{2.9}{ps}, the acting $V_G$ is strong enough to preserve the biradical open-shell singlet character of the wavefunction. Once having reached this NOON configuration, the MD is forced to evolve to "new" structural regions by adding repulsive Gaussians to $V_N$ at the current positions. Therefore, the whole landscape of $\phi_1$ and $\phi_2$ can be explored (see Fig.~S1 in the Supporting Information) while the value of both $n_H$ and $n_L$ is kept close to 1. When this degeneracy is lifted after \SI{4.5}{ps}, $V_G$ is again updated, so that the bias towards large values of $\Delta n_{meta}$ is intensified and thus, the system returns to the closed gap configuration after very short times. 

The negative of the resulting $V_N$ represents a hypersurface bearing minima with configurations of $\phi_1$ and $\phi_2$ corresponding to energetically favorable biradical structures. The surface depicted in Fig.~\ref{fig:xylylene}d has been obtained after \SI{12.0}{ps} from the sum of 110 Gaussians and subsequent symmetrization according to the equivalent nature of $\phi_1$ and $\phi_2$ as well as positive and negative values of torsion angles. It can be clearly seen that rotation of a single \ce{CH2}-moiety is the energetically most affordable way to obtain biradical character in \textit{p}-xylylene. A biradical conformation of both methylenes standing perpendicular to the central phenyl ring, however, is about \SI{15}{eV} less stable in energy, which is probably due to the two strongly localized unpaired electrons on the methylene groups, while in structures with only one rotated \ce{CH2} group, the unpaired electron of the other \ce{CH2} moiety is partly delocalized over the whole remaining $\pi$ system (see Fig.~S2 in the Supporting Information).

\subsection{[8]Annulene}

\begin{figure}
\begin{centering}
\includegraphics{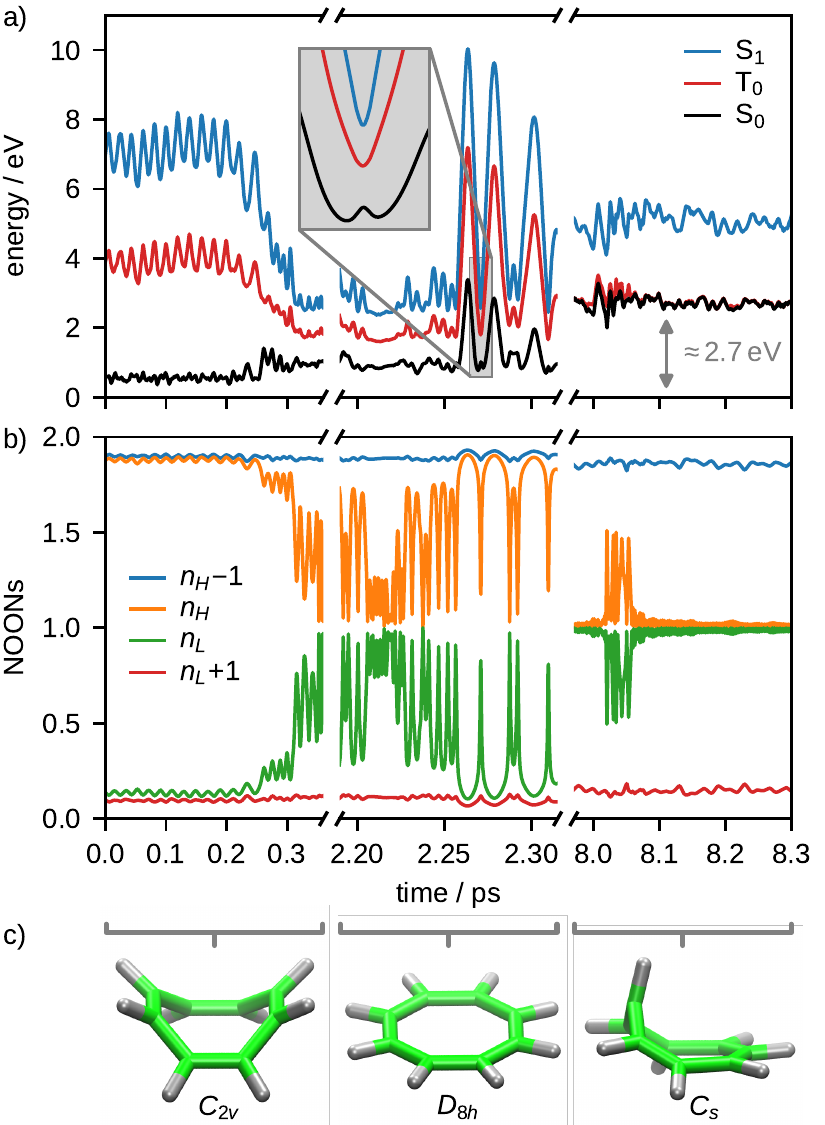}
\par\end{centering}
\protect\caption{a) Energies of the lowest singlet and triplet as well as the first excited singlet electronic states along an ASQPM trajectory of [8]annulene. b) Evolution of NOONs within the ASQPM simulation.\label{fig:annulene}}
\end{figure}

In the second example, we elucidate the biradical landscape of [8]annulene, representing a class of molecules bearing a high structural flexibility and therefore several isomeric conformations.\cite{Gellini2010} For this purpose, we employ ASQPM simulations along with the NOON gap as electronic CV and the three-dimensional Wiener number\cite{Bogdanov1989} $W$ as a versatile CV $s_1$ in Eq.~\ref{eq:metadynamics_vnoon}. The Wiener number is given as the sum over all interatomic distances in the molecule and therefore correlates with molecular shape. A small Wiener number thereby refers to a relatively compact, sphere-like structure, while planar structures lead to larger values of $W$. In Fig.~\ref{fig:annulene}a the energy evolution of the singlet and triplet as well as the first excited state along the obtained trajectory is presented. Since [8]annulene is antiaromatic, the lowest ground state structure chosen as initial conformation in the simulation is non-planar, bearing $\mathrm{C_{2v}}$ symmetry (see Fig.~\ref{fig:annulene}c), and exhibits a pronounced singlet-triplet gap of more than \SI{3}{eV}. As depicted in Fig.~\ref{fig:annulene}b, with growing $V_G$ the molecule rapidly escapes from the initial closed-shell configuration of the frontier orbitals $n_H$ and $n_L$ rearranging in a planar biradical conformation within 400 fs. 
The following \SI{7.0}{ps} are characterized by an oscillation around the $\mathrm{D_{8h}}$ symmetric structure in which HOMO and LUMO are degenerate in energy and thus the open-shell singlet configuration is favored.
As shown in the enlarged view given as inset in Fig.~\ref{fig:annulene}a, the energy gap between $S_0$ and $S_1$ decreases and increases periodically. Furthermore, the occupation of  $n_H$ and $n_L$  strongly oscillates indicating that already only slight nuclear motion of the molecule immediately cancels the biradical character of the conformer. These observation can be assigned to the presence of a Jahn-Teller conical intersection arising in axial biradicals like [8]annulene in $\mathrm{D_{8h}}$ symmetry \cite{Bonacic-Koutecky1987}. In this region of the trajectory, the molecule stays almost planar, but \ce{C-C} bond stretching leads to periodic transient formation of equal bond length and an alternating single / double bond character which breaks the symmetry. In the third section of the trajectory, $V_N$ drives the dynamics towards a $\mathrm{C_s}$ symmetric structure, baring a smaller Wiener number compared to the planar structure (see Fig.~S4 in the Supporting Information). Remarkably, from \SI{7.0}{ps} on, $\Delta n$ stays close to zero. This indicates that the electronic configuration in this range is less sensitive to structural changes. As expected, singlet and triplet energies are degenerate which allows us to characterize the structure to be of biradical character.

\begin{figure*}
\begin{centering}
\includegraphics{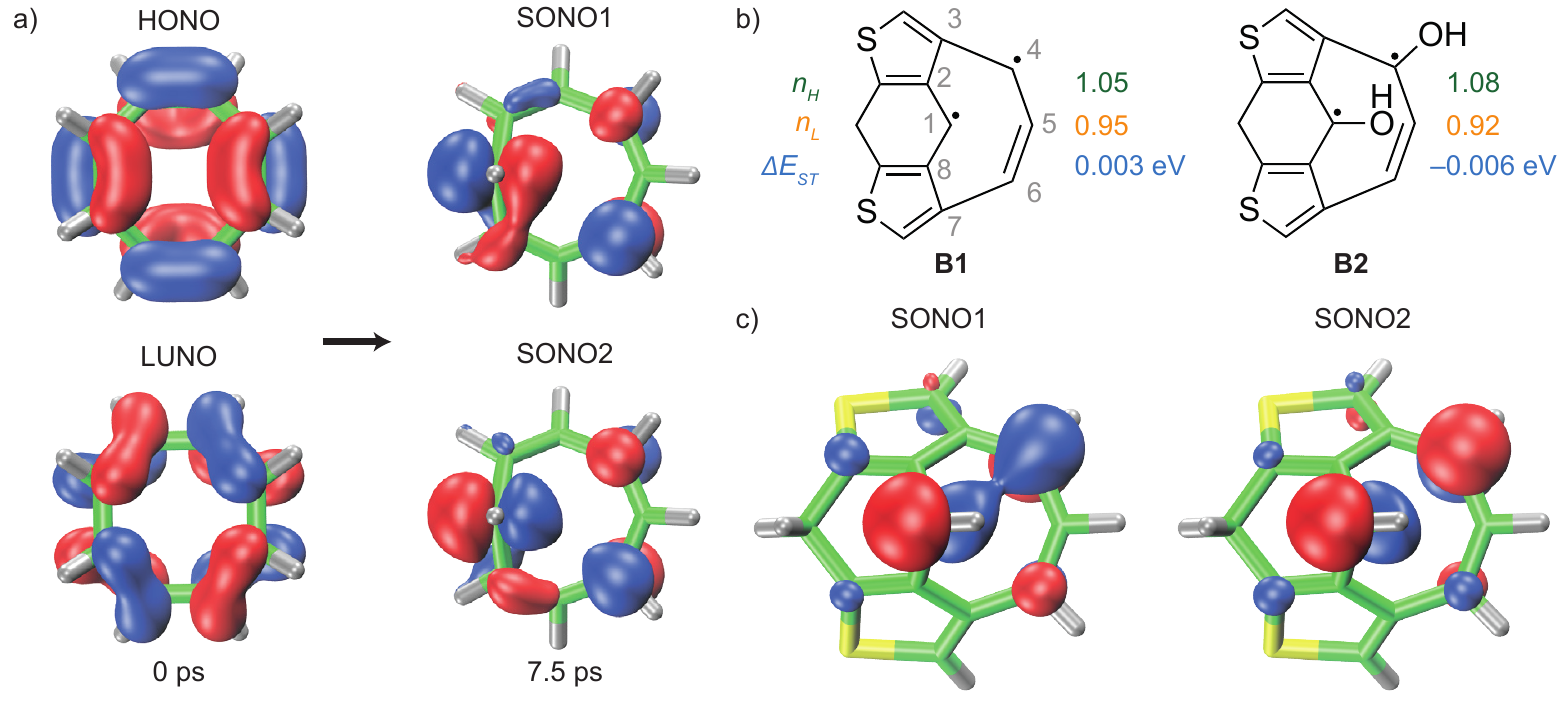}
\par\end{centering}
\protect\caption{a) HONO and LUNO of the starting structure in the trajectory from Fig.~\ref{fig:annulene} compared to the two SONOs after \SI{7.5}{ps}. b) Rationally designed substituted [8]annulene derivatives \textbf{B1} and \textbf{B2} with frontier NOONs and singlet triplet gap $\Delta E_{ST}$. The employed numbering scheme for the carbon atoms is depicted in gray. c) Optimized structure and singly-occupied NOs of the stabilized biradical. \textbf{B1}.\label{fig:annulene_stab}}
\end{figure*}

The obtained $\mathrm{C_s}$ structure is \SI{2.7}{eV} higher in energy than the ground state equilibrium structure and does not represent a minimum on the potential energy surface. By analyzing the shape of the calculated singly occupied natural orbital (SONO), the location of the respective unpaired electron can be determined. From the orbitals in Fig.~\ref{fig:annulene_stab}a, it can be seen that both HONO and LUNO are delocalized over all carbon atoms in the beginning of the trajectory. By reaching the biradical $\mathrm{C_s}$ structure, they transform into two SONOs that have the largest coefficients at C1, C4 and C6 (see Fig.~\ref{fig:annulene_stab}b for atom numbering). This allows the interpretation of two unpaired electrons each being localized at one of these three carbon atoms with high probability. This information can be used for the rational stabilization of the biradical structure by the introduction of appropriate substituents (see Fig.~\ref{fig:annulene_stab} b). First, we make use of the idea of strengthening the $\pi$ electron conjugation and thus aromatic structure by introduction of two thiophene units fused to the annulene ring allowing the electronic isolation of two double bonds adjacent to the puckered C1 atom because of the aromaticity of the two 6-electron systems. Additional sterical stability is created by another methylene bridge between the two thiophenes. By  these modifications, the $\pi$ system of annulene can be characterized as two isolated radicals separated by two closed-shell units integrated into aromatic subsystems. The resulting molecule \textbf{B1} has a distorted structure with cancelled symmetry due to partial double bond character between two of the carbon atoms in the remaining planar part. The SONOs of \textbf{B1}, given in Fig.~\ref{fig:annulene_stab} c), consequently are more localized to one side compared to the unsubstituted annulene. Furthermore, no significant contributions are observed from the carbon atoms fused to the thiophenes. On the other hand, the unpaired electron density is partly delocalized into the thiophene rings. Frontier NOONs of 1.05 and 0.95 and a singlet-triplet gap close to zero indicate an almost perfect biradical character.  The introduction of additional hydroxy groups at position 1 and 4  leads to a molecule \textbf{B2} (see Fig.~\ref{fig:annulene_stab} b) in which the biradical character is mostly preserved and which might be synthetically accessible upon preparation by twofold reduction of the corresponding closed-shell dicarbonyl species. The corresponding frontier natural orbitals are depicted in Fig.~S5 of the Supporting Information.

\section{Computational Details}
The electronic structures of both \textit{p}-xylylene and [8]annulene have been calculated on the CASSCF(8,8)/6-31G \cite{Ditchfield1971,Hehre1972,Gordon1982,Francl1982} level as implemented in Molpro2012 \cite{Werner2011}, including all $\pi$ orbitals in the active space.
MD simulations have been performed using the metaFALCON program \cite{Lindner2019} utilizing the velocity Verlet algorithm \cite{Swope1982} for the integration of the Newtonian equations of motion in steps of \SI{0.25}{fs} and temperature was controlled using a Berendsen thermostat \cite{Berendsen_1984} at \SI{300}{K}. Gaussians with a width of \SI{0.1}{} and height of \SI{0.2}{eV} have been added to $V_G$ every 200 time steps according to Eq.~\ref{eq:metadynamics}, while a value of \SI{0.1}{} has been used for $\epsilon$. $V_N$ has been composed from Gaussians with parameters of \SI{0.5}{eV} for $w$ and \ang{20} for $\delta \phi_1$ and $\delta \phi_2$ in \textit{p}-xylylene and \SI{0.1}{eV} for $w$ and \SI{0.1}{\mathring{A}} for $\delta W$ in [8]annulene. Parameters of $k=1.0$, $\epsilon_{L+1}=0.2$ and $\epsilon_{H-1}=1.8$ have been used for the construction of $V_{wall}^{L+1}$ and $V_{wall}^{H-1}$.

\section{Conclusion}
In conclusion, we have introduced a new methodology that allows for an automatic sampling of quantum property manifolds (ASQPM) giving rise to functional landscapes of molecules in analogy to the potential energy surfaces of BO states. This has been achieved by performing biased molecular dynamics simulations, that employ quantum mechanical properties derived from an electronic wavefunction as collective variable. 

As a first application we employ the ASQPM simulations in the framework of CASSCF, chosing NOONs as an electronic CV. By inserting an additional off-diagonal bias into the sub-block of the density matrix $\mathbf{D_{diag}}$ containing the HONO and LUNO NOONS $n_{H}$ and $n_{L}$ and subsequent diagonalization at the already explored regions,
we force the dynamics to systematically sample the biradical conformational space. This allows us to obtain the "biradicality landscape" of \textit{p}-xylylene. Furthermore, we use the ASQPM algorithm in order to generate a biradical conformation of [8]Annulene that could be successfully stabilized upon rational substitution as proven by structure optimization and analysis of the obtained NOONs.
We have demonstrated that information obtained from ASQPM simulations provides general insight in the structure-function relation in molecular systems and allows designing new stable biradical molecules upon rational substitution. In this way, our method should support experimental chemists in the realization of new molecules with tailored functionality. 
The method has been implemented into the metaFALCON program package \cite{Lindner2019}. 
In principle, within the ASQPM framework any function of interest that can be assigned to a property of the electronic wavefunction, both in ground and in excited state, can be explored and the simulations can be carried out in the framework of an arbitrary quantum chemical method that provides energy gradients and (if desired) excited state energies. Our future work will be devoted to to the implementation of further electronic CVs as well as the extension to QM/MM simulations allowing us to study and design catalytic reaction routes as well as new molecules of tailored functionality.

\section*{Supplementary Material}
The time evolution of the employed CVs as well as natural orbitals of \textit{p}-xylylene and \textbf{B2} and cartesian coordinates of the optimized structures are available in the Supporting Information.

\begin{acknowledgments}
J.O.L. is grateful to the Deutsche Forschungsgemeinschaft (DFG) for financial support through GRK2112:
"Molecular Biradicals: Structure, Properties and Reactivity". J.O.L. and M.I.S.R. wish to thank Roland Mitri{\'c} for fruitful discussions.
\end{acknowledgments}

\bibliography{references}

\end{document}


\newpage 

\begin{figure}[H]
\begin{centering}
\includegraphics{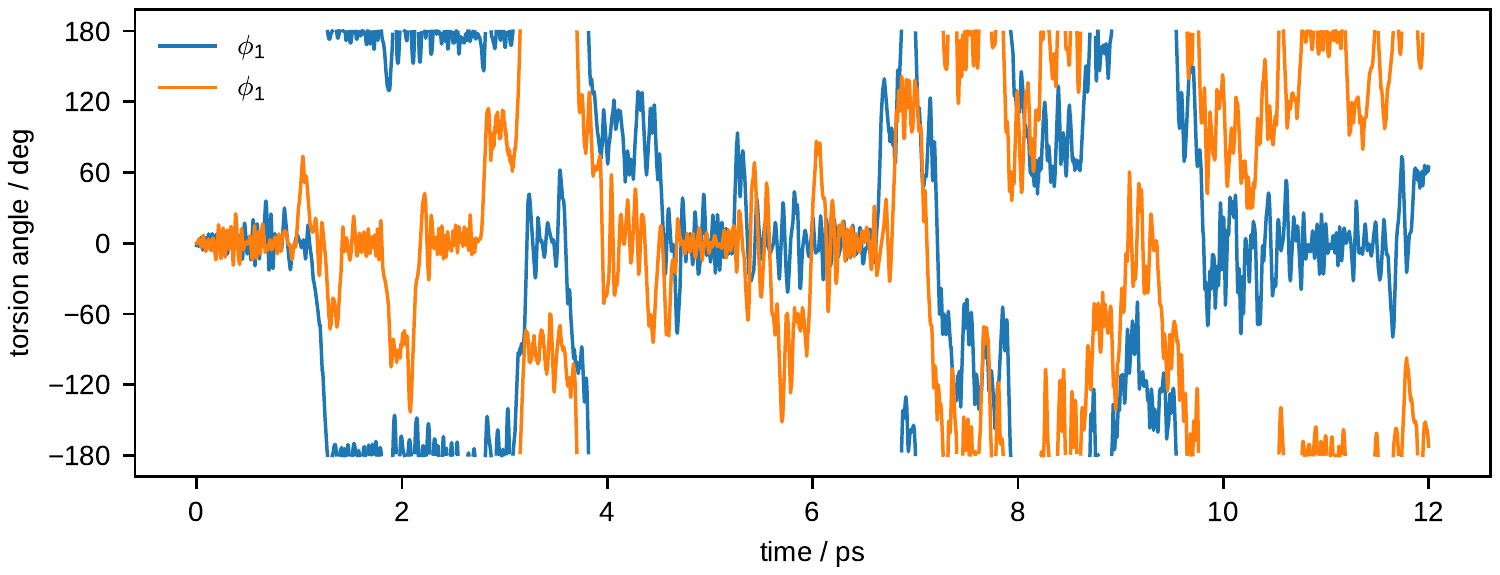}
\par\end{centering}
\caption{Torsion angles $\phi_{1}$ and $\phi_{2}$ along the ASQPM trajectory
of \textit{p}-xylylene.}
\end{figure}

\begin{figure}[H]
\begin{centering}
\subfloat[1.208 ps]{\begin{centering}
\includegraphics{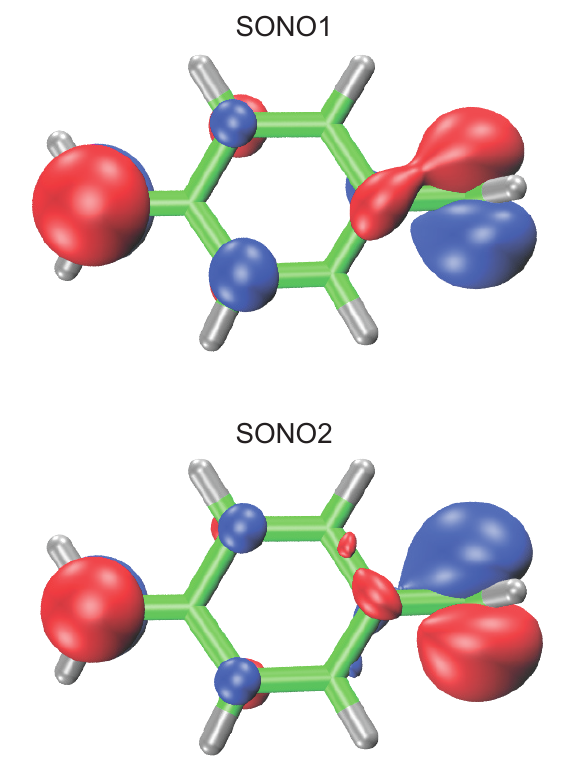}
\par\end{centering}

}\subfloat[3.990 ps]{\begin{centering}
\includegraphics{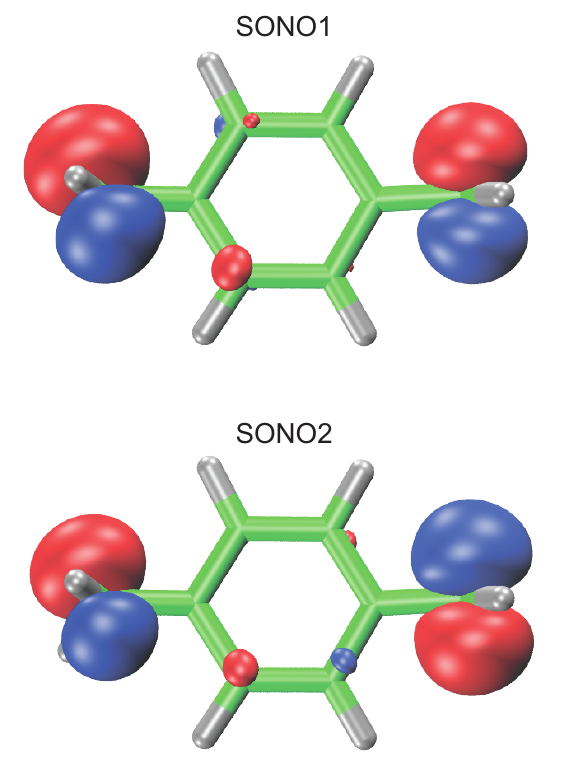}
\par\end{centering}

}
\par\end{centering}
\caption{SONOs of \textit{p}-xylylene at different time steps with a) one
rotated methylene unit and b) two rotated methylene units.}

\end{figure}

\begin{figure}[H]
\begin{centering}
\includegraphics{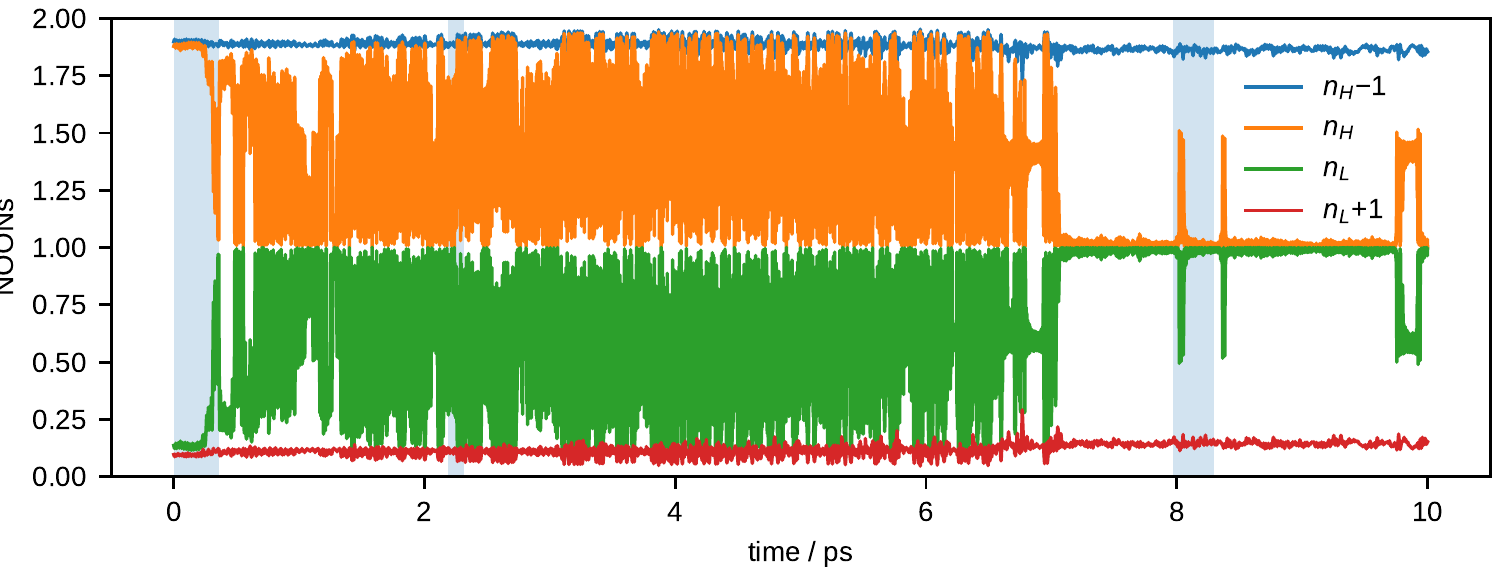}
\par\end{centering}
\caption{NOONs along the ASQPM trajectory of {[}8{]}annulene. The regions shown in Fig.~3
of the main article are highlighted with blue background.}
\end{figure}

\begin{figure}[H]
\begin{centering}
\subfloat[]{\begin{centering}
\includegraphics{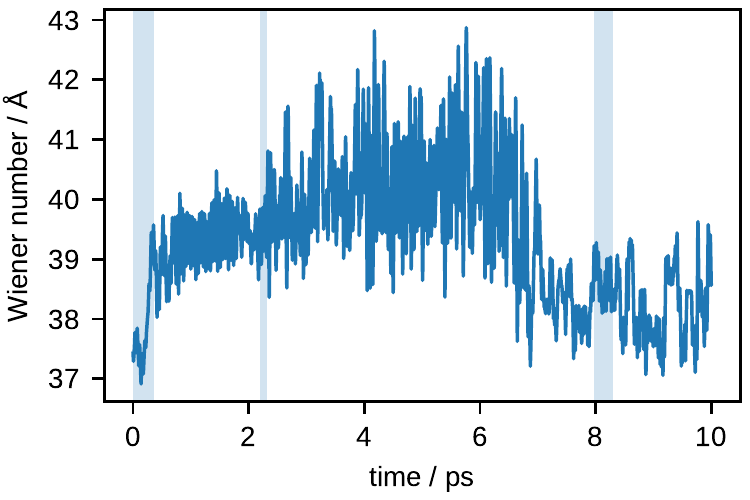}
\par\end{centering}
}\subfloat[]{\begin{centering}
\includegraphics{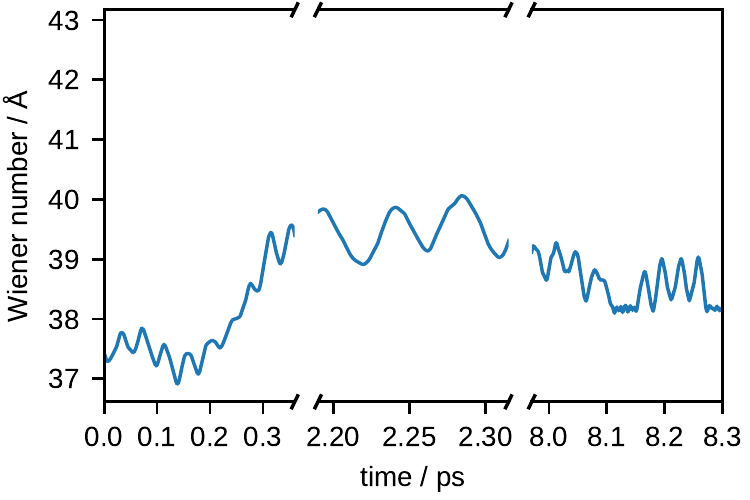}
\par\end{centering}

}
\par\end{centering}
\caption{Wiener number along the ASQPM trajectory of {[}8{]}annulene. a) Complete
time evolution of the 10 ps trajectory. The regions shown in Fig.~3
of the main article are highlighted with blue background. b) Zoomed
to the representative regions shown in Fig.~3 of the main article.}

\end{figure}

\begin{figure}[H]
\begin{centering}
\includegraphics{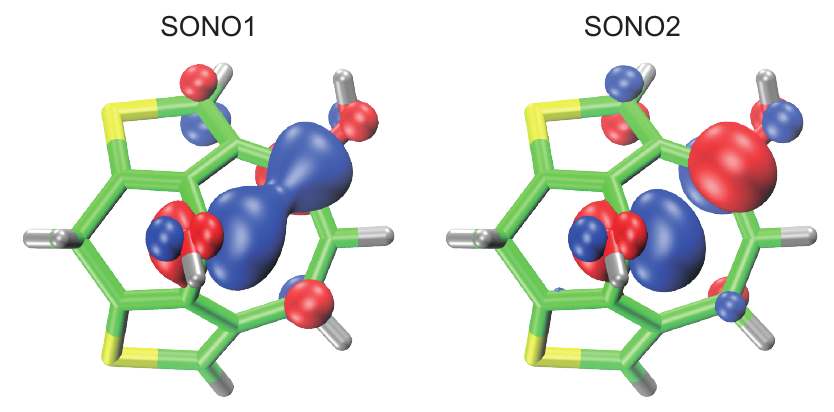}
\par\end{centering}
\caption{Optimized structure and SONOs of the stabilized biradical \textbf{B2}.}

\end{figure}

\begin{table}[H]
\caption{Cartesian coordinates of the optimized geometry of {[}8{]}annulene
in $\mathrm{{\mathring{{A}}}}$ on the CASSCF(8, 8)/6-31G level of
theory ($E=-307.53180233\,\mathrm{a.u.}$).}

\centering{}%
\begin{tabular}{cr@{\extracolsep{0pt}.}lr@{\extracolsep{0pt}.}lr@{\extracolsep{0pt}.}l}
C & -1&5830145777 & 0&2808515314 & 0&4304507653\tabularnewline
C & -1&2921523200 & -1&0354077273 & 0&4043890949\tabularnewline
C & -0&2469717743 & -1&6998044437 & -0&4021610608\tabularnewline
C & 1&0693053815 & -1&4086073491 & -0&4237177604\tabularnewline
C & 1&7610559362 & -0&3594075741 & 0&3538525632\tabularnewline
C & 1&4701486624 & 0&9568843864 & 0&3790355063\tabularnewline
C & 0&3940685061 & 1&6444599919 & -0&3647949833\tabularnewline
C & -0&9222504678 & 1&3534816475 & -0&3421442385\tabularnewline
H & -2&4192584592 & 0&5985718360 & 1&0310041868\tabularnewline
H & -1&9110653678 & -1&6981959869 & 0&9861516444\tabularnewline
H & -0&5853137206 & -2&5390744312 & -0&9868922547\tabularnewline
H & 1&7114187327 & -2&0306067013 & -1&0249579002\tabularnewline
H & 2&6206437139 & -0&6944034225 & 0&9102626244\tabularnewline
H & 2&1124850714 & 1&6021790304 & 0&9548877056\tabularnewline
H & 0&7091754800 & 2&5011436933 & -0&9371550193\tabularnewline
H & -1&5875112669 & 1&9926022891 & -0&8986628435\tabularnewline
\end{tabular}
\end{table}

\begin{table}[H]
\caption{Cartesian coordinates of the optimized geometry of \textbf{B1} in
$\mathrm{{\mathring{{A}}}}$ on the CASSCF(8, 8)/6-31G level of theory
($E=-1291.58989906\,\mathrm{a.u.}$).}

\centering{}%
\begin{tabular}{cr@{\extracolsep{0pt}.}lr@{\extracolsep{0pt}.}lr@{\extracolsep{0pt}.}l}
C & -1&4906566719 & 0&3543665167 & 0&7460648330\tabularnewline
C & -1&7759335667 & -0&7004577140 & -0&2249453588\tabularnewline
C & -0&6869022731 & -1&5210095772 & -0&7324828926\tabularnewline
C & 0&6925999274 & -1&3772860520 & -0&1988337635\tabularnewline
C & 1&4047877325 & -0&2572350908 & 0&0693296858\tabularnewline
C & 1&2182773240 & 1&1942049259 & -0&0530776426\tabularnewline
C & 0&0300038559 & 1&9429811645 & -0&4967878064\tabularnewline
C & -1&3046502013 & 1&5505121967 & -0&0716572361\tabularnewline
C & -1&0282257111 & -2&1476750885 & -1&8938835242\tabularnewline
H & 2&4011113576 & -0&4705204675 & 0&4180948726\tabularnewline
C & -0&0118851705 & 2&8241123716 & -1&5139900914\tabularnewline
C & -2&3137747376 & 1&8919830171 & -0&9201030568\tabularnewline
C & -3&4655589287 & 0&8991044472 & -1&1184928961\tabularnewline
H & -4&1908089058 & 0&9907867214 & -0&3175471307\tabularnewline
H & -3&9820016321 & 1&0732290731 & -2&0501588609\tabularnewline
C & -2&8062020109 & -0&4861988163 & -1&0931129797\tabularnewline
H & -0&9376173319 & 0&1786415082 & 1&6447150876\tabularnewline
H & 2&1317588720 & 1&7534536840 & 0&0310470387\tabularnewline
H & 1&2216114441 & -2&3012578569 & -0&0456018864\tabularnewline
S & -2&6617091111 & -1&6261468899 & -2&5048344020\tabularnewline
S & -1&6917493399 & 3&0896606711 & -2&1371453970\tabularnewline
H & 0&8055996290 & 3&2840623646 & -2&0243518321\tabularnewline
H & -0&4311345499 & -2&7937911090 & -2&4988647604\tabularnewline
\end{tabular}
\end{table}

\begin{table}[H]
\caption{Cartesian coordinates of the optimized geometry of \textbf{B2} in
$\mathrm{{\mathring{{A}}}}$ on the CASSCF(8, 8)/6-31G level of theory
($E=-1441.22883733\,\mathrm{a.u.}$).}

\centering{}%
\begin{tabular}{cr@{\extracolsep{0pt}.}lr@{\extracolsep{0pt}.}lr@{\extracolsep{0pt}.}l}
C & -1&4766746189 & 0&3402677967 & 0&7625665170\tabularnewline
C & -1&7484442231 & -0&7467223757 & -0&1989843396\tabularnewline
C & -0&6495247606 & -1&5104751013 & -0&7573204184\tabularnewline
C & 0&7407446498 & -1&3638601257 & -0&2528862428\tabularnewline
C & 1&4460524301 & -0&2467674314 & 0&0145610072\tabularnewline
C & 1&2270746695 & 1&2012858132 & -0&0715947021\tabularnewline
C & 0&0473141569 & 1&9424014808 & -0&5131407413\tabularnewline
C & -1&2686389461 & 1&5012396091 & -0&1154383734\tabularnewline
C & -1&0150089983 & -2&1353929916 & -1&9128964328\tabularnewline
H & 2&4574120699 & -0&4217576674 & 0&3329647233\tabularnewline
C & -0&0203339531 & 2&8673601963 & -1&5214538591\tabularnewline
C & -2&2713095319 & 1&8216883809 & -0&9370113014\tabularnewline
C & -3&4566586761 & 0&8552302102 & -1&0634969143\tabularnewline
H & -4&1476529304 & 0&9876615054 & -0&2401256141\tabularnewline
H & -3&9967239592 & 1&0118228305 & -1&9843997843\tabularnewline
C & -2&8198692891 & -0&5484151785 & -1&0247921144\tabularnewline
O & -2&3121498065 & 0&5387479034 & 1&8469703072\tabularnewline
O & 2&4400315740 & 1&8765565845 & -0&0787223334\tabularnewline
H & 1&2795355254 & -2&2849225005 & -0&1218412745\tabularnewline
S & -2&6857927785 & -1&6615267062 & -2&4574403787\tabularnewline
S & -1&7237591255 & 3&0767153558 & -2&1210732155\tabularnewline
H & 0&7768566026 & 3&3471603570 & -2&0438540914\tabularnewline
H & -0&4296766157 & -2&7754797910 & -2&5351693257\tabularnewline
H & -2&5098668971 & -0&2569012871 & 2&3294581439\tabularnewline
H & 2&3728234320 & 2&7983431327 & 0&1435507586\tabularnewline
\end{tabular}
\end{table}